\documentclass[%
reprint,
 amsmath,amssymb,
 aps,
 prl,
]{revtex4-2}

\usepackage{graphicx}
\usepackage{dcolumn}
\usepackage{bm}
\usepackage{xcolor}

\usepackage[normalem]{ulem}

\definecolor{fpgreen}{rgb}{0,.5,0}

\newcommand{\R}{\mathbb{R}}
\newcommand{\alphab}{\boldsymbol\alpha}
\newcommand{\cT}{\mathcal{T}}
\newcommand{\Gr}{\mathcal{G}r}
\DeclareMathOperator{\Log}{Log}
\DeclareMathOperator{\Exp}{Exp}
\DeclareMathOperator{\Tr}{Tr}

\begin{document}

\preprint{APS/123-QED}

\title{A Quasi Time-Reversible scheme based on density matrix extrapolation on the Grassmann manifold for Born-Oppenheimer Molecular Dynamics}


\author{Federica Pes}
\affiliation{
Dipartimento di Chimica e Chimica Industriale, Universit\`a di Pisa,
Via G. Moruzzi 13, 56124 Pisa, Italy
}
\author{\'Etienne Polack}
\affiliation{
CERMICS, \'Ecole des Ponts and Inria Paris, 6 \& 8 avenue Blaise Pascal, 77455 Marne-la-Vall\'e, France
}
\author{Patrizia Mazzeo}
\affiliation{
Dipartimento di Chimica e Chimica Industriale, Universit\`a di Pisa,
Via G. Moruzzi 13, 56124 Pisa, Italy
}

\author{Genevi\`eve Dusson}
\affiliation{
Laboratoire de Math\'ematiques de Besan\c{c}on, UMR CNRS 6623,
  Universit\'e de Franche-Comt\'e, 16 route de Gray, 25030 Besan\c{c}on, France
}%
\author{Benjamin Stamm}
\affiliation{
Institute of Applied Analysis and Numerical Simulation, University of Stuttgart, 70569 Stuttgart, Germany
}
\author{Filippo Lipparini}
\email{filippo.lipparini@unipi.it}
\affiliation{
Dipartimento di Chimica e Chimica Industriale, Universit\`a di Pisa,
Via G. Moruzzi 13, 56124 Pisa, Italy
}


\date{\today}

\begin{abstract}

This article proposes a so-called Quasi Time-Reversible (QTR G-Ext) scheme based on Grassmann extrapolation of density matrices for an accurate calculation of initial guesses in Born-Oppenheimer Molecular Dynamics simulations.
The method shows excellent results on four large molecular systems, ranging from 21 to 94 atoms simulated with Kohn-Sham density functional theory surrounded with a classical environment with 6k to 16k atoms.
Namely, it clearly reduces the number of self-consistent field iterations, while keeping a similar energy drift as in the extended Lagrangian Born-Oppenheimer method.

\end{abstract}

\maketitle

Ab-initio, Born-Oppenheimer molecular dynamics (BOMD) is a very powerful and versatile tool to simulate molecular processes where the quantum nature of the system is not negligible. 
Unfortunately, this comes at a high computational price, which stems from the necessity of solving the quantum mechanical (QM) equations, typically Kohn-Sham Density Functional Theory (KS-DFT) equations, to compute the energy and forces at every time-step. 
Such equations are nonlinear and are solved using a fixed-point iterative method known as Self-Consistent Field~\cite{roothaan} (SCF). BOMD simulations, that require one to perform tens of thousands of SCF calculations, rely thus heavily on extrapolation techniques~\cite{fang_existence_2016,alfe_ab_1999,arias_ab_1992,pulay_fock_2004,herbert_accelerated_2005,hutter_exponential_1994,vandevondele_efficient_2003,vandevondele_quickstep_2005,Niklasson_PRL_TRBOMD,niklasson_extended_2008,niklasson_extended_2009,vitale2017} that, by using converged solutions from previous iterations, compute an accurate guess for the SCF, limiting thus the number of iterations required to achieve convergence. 
A significant contribution to this field was given by Niklasson and co-workers in 2006, with their work on the time-reversible extrapolation for Born-Oppenheimer Molecular Dynamics (BOMD) \cite{Niklasson_PRL_TRBOMD}. 
The core concept involves generating a guess density matrix by combining the density matrices from previous steps in a symmetric and time-reversible manner. However, numerical applications showed that enforcing an \emph{exact} time-reversibility can lead to errors accumulating in long-time simulations, 
spoiling thus the convergence properties of the algorithm in the long run. This led to the development of the Extended Lagrangian Born-Oppenheimer approach (XLBO) in 2008 \cite{niklasson_extended_2008,niklasson_extended_2009}. 
In this particular case, the time-reversible extrapolation is augmented by the inclusion of a dissipative term, which serves to reduce numerical fluctuations. 
In the last few years, the XLBO method has also been proposed in a SCF-less formulation,~\cite{Niklasson_JCP_NGXLBO,Niklasson_JCTC_NGXLBO,vitale2017} where the density computed using the XLBO procedure is used directly without further SCF iterations.


In Niklasson's XLBO scheme, the guess density is propagated in time subject to a potential that forces it to be close to the converged density. However, the guess density obtained with XLBO is not exactly idempotent~\cite{niklasson_extended_2008}, which is in practice a problem that can be either ignored, or easily addressed by, for instance, McWeeny purification~\cite{McWeeny_RMP_Purification,polack2021}. 

Recently, we proposed a different strategy to compute a guess density using linear extrapolation. This is non-trivial, because in general a linear combination of density matrices does not preserve idempotency or, in other words, density matrices belong to a differentiable manifold called Grassmann manifold and not to a vector space. 
Our approach uses tools from differential geometry to map the Grassmann manifold onto its tangent space, which is a vector space. It then performs a linear extrapolation on the tangent space, and then maps back the extrapolated density to the manifold. We named such a method Grassmann extrapolation (G-Ext)~\cite{polack2021,polack_approximation_2020}. 
G-Ext has been successfully adopted in the Pisa-group for both ground- and excited-state SCF-based BOMD simulations in a polarizable multiscale framework~\cite{Mazzeo2023}.
While G-Ext is very effective, indeed outperforming XLBO in terms of the average number of SCF iterations required to achieve convergence along a MD trajectory~\cite{polack2021}, numerical experiments have shown that the extrapolation introduces a bias causing a drift in the total energy for NVE simulations. 
Such an energy drift is modest (few kcal/mol in 10 ps), but non negligible, at least for not-too-tightly-converged calculations~\cite{polack2021}. In this contribution, we address such a limitation by introducing a new strategy to perform the extrapolation, that we name Quasi Time-Reversible Grassmann extrapolation method (QTR G-Ext). 
This approach leverages the principles of differential geometry, similarly to the previous method, but offers enhanced accuracy and speed in extrapolating the density matrix during a BOMD simulation, as well as excellent energy conservation properties.

Given a $\mathcal{N}$-dimensional basis, the SCF solves the following nonlinear eigenvalue problem which consists to find a matrix $C$ and a diagonal matrix $E$ such that
\[
\begin{cases}
F(D)C=SCE \\
C^TSC=I_N \\
D=CC^T,
\end{cases}
\]
where $C\in\R^{\mathcal{N}\times N}$ contains the $\mathcal{N}$ coefficients of the $N$ occupied molecular orbitals, $D\in\R^{\mathcal{N}\times \mathcal{N}}$ is the density matrix, $E\in\R^{N\times N}$ is a diagonal matrix which entries are the energy levels, $F$ denotes the DFT operator, $S\in\R^{\mathcal{N}\times\mathcal{N}}$ is the overlap matrix, and $I_N$ denotes the identity matrix of order $N$. 

We assume that the density matrix is orthogonal. In any case, it can be transformed into such matrix by considering the L\"owdin factorization of the overlap matrix $S$ and consequently the modified coefficient matrix $\widetilde C = S^{1/2}C$. Then the normalized density matrix $\widetilde D = \widetilde C \widetilde C^T = S^{1/2}DS^{1/2}$ belongs to the manifold
\[
\Gr(N,\mathcal{N}) = \left\{ D\in\R^{\mathcal{N}\times\mathcal{N}} | D^2=D=D^T, \Tr(D)=N \right\},
\]
which is isomorphic to the so-called ``Grassmann manifold'', therefore we identify $\Gr$ by this name.
From now on, we assume that the density matrix has been orthonormalized and we denote it by $D$.

Since $\Gr$ is a differential manifold, given a point $D_0\in\Gr$, there exists a tangent space $\cT_{D_0}\subset\R^{\mathcal{N}\times N}$, such that tangent vectors $\Gamma(D)\in\cT_{D_0}$ can be associated to nearby points $D\in\Gr$.

In MD, $t\to\bm{R}(t)$ represents the trajectory of the nuclei.
The transformation of the electronic structure can be interpreted as a trajectory denoted by $t\to D_{\bm{R}(t)}$ on the manifold.
In order not to burden the notation, we simply indicate $D$ in place of $D_{\bm{R}(t)}$.
The objective is to determine a suitable approximation for the density matrix at the next step of the molecular dynamics trajectory by extrapolating the densities from previous steps.
Since the tangent space $\cT_{D_0}$ is a vector space, we approximate the density matrix on $\cT_{D_0}$.
In order to solve the extrapolation problem, we decompose the mapping $\bm{R}\to D$ as a composition of several maps
\begin{equation}\label{maps}
\begin{aligned}
\R^{3M} & \longrightarrow \mathcal{D} \longrightarrow \mathcal{T}_{D_0} \longrightarrow \Gr\\
\bm{R} & \longmapsto d \;\; \longmapsto  \Gamma \quad \longmapsto D,
\end{aligned}
\end{equation}
where
the first function $\bm{R} \mapsto d$ is a map from atomic positions to molecular descriptors. Here, as a descriptor, we use the Coulomb matrix~\cite{Rupp2012} $d \in \mathbb{R}^{N_{\rm QM}\times N_{\rm QM}}$, 
\begin{equation}\label{coul}
(d)_{ij} = 
\begin{cases}
0.5z_i^{2.4} & i=j,\\
\dfrac{z_iz_j}{\|\bm{R}(t_i)-\bm{R}(t_j)\|} & i\neq j,
\end{cases}
\end{equation}
where $N_{\rm QM}$ is the number of atoms treated quantum mechanically and $z_i$ denotes the nuclear charge of the $i$th atom. Note that other descriptors can also be considered.
We will detail the crucial mapping $d \mapsto \Gamma$ below.
The mapping $\Gamma\mapsto\Exp(\Gamma)=D$ is the so-called Grassmann exponential which maps tangent vectors on $\cT_{D_0}$ to $\Gr$, and it is a locally bijective function in a neighborhood of $D_0$. Its inverse $D\mapsto \Log(D)=\Gamma(D)$ is the Grassmann logarithm. These mappings are computed by means of the singular value decomposition (SVD). For mathematical details, the interested reader is referred to \cite{polack_approximation_2020,zimmermann_manifold_2019,edelman_geometry_1998}.
In our method, during the MD, we use a fixed reference point $D_0$ to construct the tangent space $\cT_{D_0}$.

Let $n$ be the current time step of the MD.
Given previous $q$ snapshots $\Gamma_{n-i}=\Log(D_{n-i})$, for $i=1,\ldots,q$, the approximation of the density matrix on the tangent space is written as
\begin{equation}\label{atr}
\widetilde\Gamma_n = -\Gamma_{n-q} + 
\sum_{i=1}^{\widetilde q} \alpha_i \left( \Gamma_{n-i}+\Gamma_{n-q+i} \right),
\end{equation}
where $\widetilde q = q/2$ if $q$ is even, while $\widetilde q = (q-1)/2$ if $q$ is odd.
We remark that if in Eq. \eqref{atr}, the term $\Gamma_{n-q}$ is substituted by $\widetilde\Gamma_{n-q}$, a ``fully'' time-reversible approach (instead of quasi time-reversible) is obtained. Numerical experiments with the fully time-reversible approach, that are reported in the Supporting Information (SI), showed good behavior for total energy conservation, but unfortunately a strong increase in the number of performed SCF iterations.

The descriptors are involved in the computation of the coefficients $\alphab=[\alpha_1,\ldots,\alpha_{\widetilde q}]^T$ appearing in Eq.~\eqref{atr}.
Indeed, they are computed by solving the least-squares problem with Tikhonov regularization
\[
\min_{\alphab\in\R^{\widetilde q}} \left\{ \left\| d_n + d_{n-q} - 
\sum_{i=1}^{\widetilde q} \alpha_i \left( d_{n-i}+d_{n-q+i} \right)
\right\|^2 + \varepsilon^2 \left\| \alphab \right\|^2 \right\},
\]
where $\|\cdot\|$ denotes the $\ell^2$-norm and $\varepsilon>0$ is the regularization parameter. Since the Coulomb matrix \eqref{coul} is symmetric, in the above formula $d_j$ represents the vectorized Coulomb matrix considering the lower triangle. In matrix form, it corresponds to solving the following least-squares problem
\[
\min_{\alphab\in\R^{\widetilde q}}
\left\|
\begin{bmatrix} \widehat d \\ \bm{0} \end{bmatrix} -
\begin{bmatrix} \widehat D \\ \varepsilon I_{\widetilde q} \end{bmatrix} \alphab
\right\|^2,
\]
where the vector $\widehat d = d_n + d_{n-q}$ is padded with $\widetilde q$ zeroes,
$\widehat D\in\R^{N_d\times \widetilde q}$ is the matrix which columns are defined as $\widehat D_{\cdot,i}=d_{n-i}+d_{n-q+i}$, and $I_{\widetilde q}$ is the identity matrix of order $\widetilde q$.
Then the initial guess for the density matrix is obtained as the composition of the three maps in~\eqref{maps}, where the second map $d \mapsto \Gamma$ is given by~\eqref{atr}. Note that if this second map denoted by $f$ was linear, then the guess would be close to exact, namely
\begin{align*}
\Gamma_n & = f(d_n) \approx
f\left( - d_{n-q} + 
\sum_{i=1}^{\widetilde q} \alpha_i \left( d_{n-i}+d_{n-q+i} \right) \right)  \\
& = - f\left(d_{n-q}\right) + 
\sum_{i=1}^{\widetilde q} \alpha_i \left[ f\left(d_{n-i}\right) + f\left(d_{n-q+i}\right) \right] \\
& = -\Gamma_{n-q} + 
\sum_{i=1}^{\widetilde q} \alpha_i \left( \Gamma_{n-i}+\Gamma_{n-q+i} \right) = \widetilde\Gamma_n.
\end{align*}

The number $q$ of density matrices taken at previous steps and the value of the regularization parameter $\varepsilon$ are chosen in a heuristic manner: we computed the error $\|\Gamma_n - \widetilde{\Gamma}_n \|$ for different values of $q$ and $\varepsilon$, specifically $q=3,4,\ldots,20$ and $\varepsilon = 0.001, 0.002, 0.005, 0.01, 0.02, 0.05$, and we selected the combination $(q,\varepsilon)$ corresponding to the minimal error.
When the SCF convergence threshold is $10^{-5}$, we found that good values are $q=5$ and $\varepsilon=0.005$, while if it is fixed to $10^{-7}$, we found $q=4$ and $\varepsilon=0.001,0.002$. Additional details on the selection of $q$ and $\varepsilon$ values can be found in Section S1 of the SI.

The QTR G-Ext approach is tested on four different systems.
The first system is dimethylaminobenzonitrile (DMABN) in methanol. The second system is 3-hydroxyflavone (3HF) in acetonitrile. The last two systems (OCP and AppA) are chromophores embedded in a biological matrix-namely, a carotenoid in the orange carotenoid protein (OCP) and a flavin in the AppA Blue-Light Using Flavin photoreceptor \cite{Bondanza_Chem_OCP, Bondanza_JACS_OCP,Macaluso_CS_AppA}.
Some information on the systems is reported in Table \ref{tab:size}.
\begin{table}
\caption{\label{tab:size} Summary of systems' size: number of QM atoms $N_{\text{QM}}$, number of MM atoms $N_{\text{MM}}$, number of QM basis functions $\mathcal{N}$, number of occupied orbitals $N$, and size of descriptors $N_d$.}
\begin{ruledtabular}
\begin{tabular}{lrrrrr}
System & $N_{\text{QM}}$ & $N_{\text{MM}}$ & $\mathcal{N}$ & $N$ & $N_d$\\ \hline
DMABN & 21 & 6843 & 185 & 39 & 234\\
3HF & 28 & 15046 & 290 & 62 & 409\\
AppA & 31 & 16449 & 309 & 67 & 468\\
OCP & 94 & 6058 & 734 & 154 & 4468\\
\end{tabular}
\end{ruledtabular}
\end{table}
KS-DFT has been adopted to describe the QM subsystem, with the B3LYP hybrid functional \cite{B3LYP} and the 6-31G(d) Pople's basis set \cite{Hehre1972}. This is coupled with a polarizable description of the environment, using the AMOEBA forcefield \cite{ponder2010}.
For each system, we performed a QM/AMOEBA geometry optimization until a root-mean-square norm on the forces of 4 kcal/mol/Å is found and finally a 2 ps QM/AMOEBA NVT equilibration to obtain the starting point of the simulations presented in this work. 

All simulations have been performed using the Gaussian-Tinker interface \cite{Nottoli2020,Loco2017,Loco_CS_PB,gdv}. 
We implemented the QTR G-Ext extrapolation approach in Tinker \cite{Tinker,Tinker-HP}.

To assess the quality of the guess density obtained by the QTR G-Ext extrapolation, we performed 10 ps BOMD simulations, with 0.5 fs time step, in the NVE ensemble, using the velocity Verlet integrator \cite{verlet}. All systems were tested with an SCF convergence threshold fixed to $10^{-5}$ and $10^{-7}$ with respect to the RMS variation of density. 
We compare our approach in terms of energy stability and number of iterations required to reach convergence with other two extrapolation schemes, which are the G-Ext scheme \cite{polack2021}
\begin{equation*}
\widetilde\Gamma_n =  
\sum_{i=1}^{q} \alpha_i \Gamma_{n-i}, \qquad q=6,
\end{equation*}
where the $\alpha_i$ are computed by solving
\[
\min_{\alphab\in\R^{q}} \left\{ \left\| d_n - \sum_{i=1}^{q} \alpha_i d_{n-i} \right\|^2 + \varepsilon^2 \left\| \alphab \right\|^2 \right\},\quad \varepsilon=0.01,
\]
and XLBO without McWeeny purification \cite{niklasson_extended_2008,niklasson_extended_2009}
\begin{equation*}
\widetilde{D}_n = 2 \widetilde{D}_{n-1} - \widetilde{D}_{n-2} + 
\kappa \left( D_{n-1} - \widetilde{D}_{n-1} \right) +
c \sum_{i=1}^{8} \alpha_i \widetilde{D}_{n-i},
\end{equation*}
with fixed parameters $\kappa=1.86$, $c=0.0016$, and $\alphab=(-36,99, -88, 11, 32, -25, 8, -1)$.

\begin{figure}
\includegraphics[width=\linewidth]{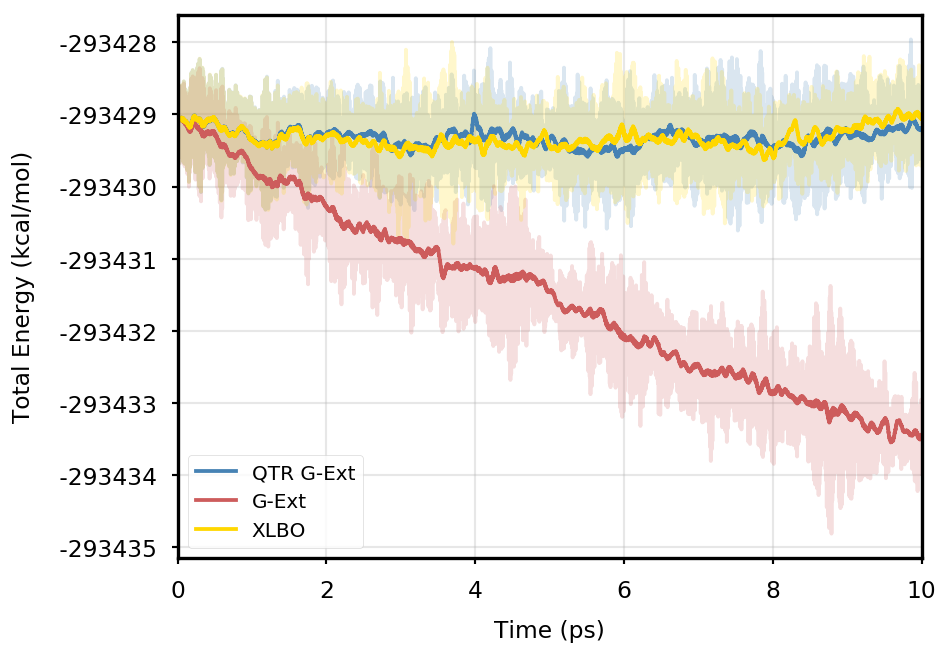}
\caption{\label{fig5:energy_dmabn} Total energy as a function of simulation time for DMABN, using a $10^{-5}$ convergence threshold for the SCF.}
\end{figure}
\begin{figure}
\includegraphics[width=\linewidth]{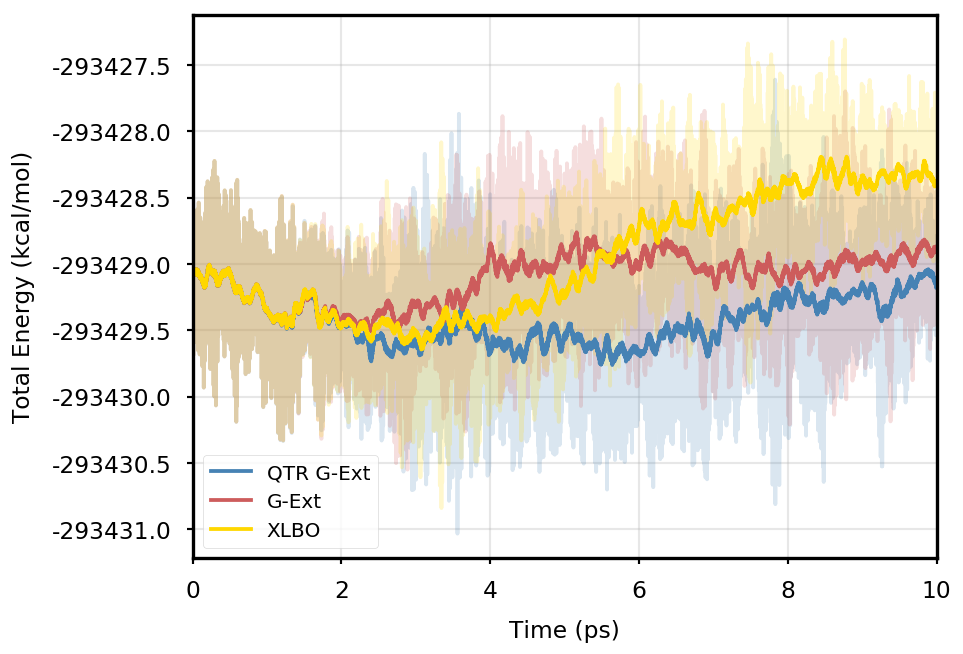}
\caption{\label{fig7:energy_dmabn} Total energy as a function of simulation time for DMABN, using a convergence threshold for the SCF of $10^{-7}$.}
\end{figure}

Figure \ref{fig5:energy_dmabn} provides the plot of the total energy along the DMABN simulation, with a $10^{-5}$ SCF convergence threshold. Despite the non-fully time-reversible formulation of our newly implemented approach, we observe a great improvement with respect to the G-Ext scheme.
In particular, the QTR G-Ext method resembles the fully time-reversible scheme XLBO. 
The same behaviour is almost imperceptible when the SCF convergence is set to $10^{-7}$ (Figure \ref{fig7:energy_dmabn}), since the accumulation of errors that generates the energy drift when G-Ext is used is lower, so we can appreciate the same trend with all the extrapolation schemes. Analogous figures are reported in Section S2 of the SI for all tested systems. 
To better evaluate the energy stability, we consider the average short-time fluctuation (STF) of the energy, which is computed by getting the RMS of the energy fluctuation every 50 fs and averaging over the trajectory, 
and the long-time drift (LTD) for a long-time analysis, that is the slope of the linear regression line of the energy. Tables \ref{tab:stf-ltd-5} and \ref{tab:stf-ltd-7} disclose STF and LTD for convergence thresholds $10^{-5}$ and $10^{-7}$, respectively.  
QTR G-Ext, G-Ext, and XLBO show comparable STF, which is specific for the system and is related to the time step for the integration. On the other hand, the absolute value of LTD is in general higher for $10^{-5}$ simulations, in particular for G-Ext. We can state that the QTR G-Ext method solves the energy-drift issue of G-Ext, showing an LTD that is always similar to the XLBO one, suggesting again a good time-reversible behaviour.

\begin{table}
\caption{\label{tab:stf-ltd-5} Short- and Long-Time Stability Analysis of the QTR G-Ext, G-Ext, and XLBO methods. SCF convergence threshold $10^{-5}$.}
\begin{ruledtabular}
\begin{tabular}{lcccccccc}
      & \multicolumn{2}{c}{DMABN} & \multicolumn{2}{c}{3HF} & \multicolumn{2}{c}{AppA} & \multicolumn{2}{c}{OCP} \\
      & STF & LTD & STF & LTD & STF & LTD & STF & LTD \\ \hline
QTR G-Ext   & 0.33 & -0.01 & 0.62 & -0.40 & 0.57 & -0.08 & 0.36 & -0.23   \\
G-Ext & 0.35 & -0.43 & 0.61 & -0.94 & 0.56 & -0.93 & 0.38 & -1.38   \\
XLBO  & 0.32 &  0.01 & 0.57 & -0.42 & 0.59 & 0.14 & 0.39 & -0.28   \\ 
\end{tabular}
\end{ruledtabular}
\end{table}

\begin{table}
\caption{\label{tab:stf-ltd-7} Short- and Long-Time Stability Analysis of the QTR G-Ext, G-Ext, and XLBO methods. SCF convergence threshold $10^{-7}$.}
\begin{ruledtabular}
\begin{tabular}{lcccccccc}
      & \multicolumn{2}{c}{DMABN} & \multicolumn{2}{c}{3HF} & \multicolumn{2}{c}{AppA} & \multicolumn{2}{c}{OCP} \\
      & STF & LTD & STF & LTD & STF & LTD & STF & LTD \\ \hline
QTR G-Ext  & 0.37 & 0.01 & 0.59 & -0.30 & 0.53 & 0.18 & 0.38 & -0.16 \\
G-Ext & 0.33 & 0.04 & 0.60 & -0.27 & 0.54 & 0.06 & 0.38 & -0.20 \\
XLBO  & 0.32 & 0.13 & 0.64 & -0.37 & 0.56 & 0.06 & 0.38 & -0.08 \\ 
\end{tabular}
\end{ruledtabular}
\end{table}

The gain of our new methodology is not only in terms of accuracy (energy stability), but also in terms of the computational time of the simulation. Tables \ref{tab:scf5} and \ref{tab:scf7} report the average number of SCF iterations required to achieve convergence, as well as the standard deviation for $10^{-5}$ and $10^{-7}$ SCF thresholds, respectively. We remark that each strategy requires $q$ previous density matrices, before having them available a standard SCF is performed. Therefore, for the computation of average and standard deviation, we discard the first $q$ points.
The two tables show that for all the tested systems, the QTR G-Ext method requires the lowest number of SCF iterations, for both convergence thresholds. Moving averages of SCF iteration numbers during the simulations for all systems and with both SCF convergence thresholds are reported in Section S3 of the SI.

\begin{table}
\caption{\label{tab:scf5} Performance of the QTR G-Ext method compared with the G-Ext method and XLBO algorithm. Average $\overline{k}$ and standard deviation $\sigma$ of SCF iterations. Convergence threshold $10^{-5}$.}
\begin{ruledtabular}
\begin{tabular}{lcccccccc}
      & \multicolumn{2}{c}{DMABN} & \multicolumn{2}{c}{3HF} & \multicolumn{2}{c}{AppA} & \multicolumn{2}{c}{OCP} \\
      & $\overline{k}$ & $\sigma$ & $\overline{k}$ & $\sigma$ & $\overline{k}$ & $\sigma$ & $\overline{k}$ & $\sigma$ \\ \hline
QTR G-Ext  &  3.04   &  0.22  &  2.98   &  0.21  & 3.00  & 0.02  &  2.96   &  0.31  \\
G-Ext &  3.55   &  0.85  &  3.16   &  0.69  & 3.03 & 0.54 &  2.91   &  0.41  \\
XLBO  &  4.00   &  0.05  &  4.00   &  0.00  & 4.00 & 0.07 &  4.00   &  0.01      \\ 
\end{tabular}
\end{ruledtabular}
\end{table}

\begin{table}
\caption{\label{tab:scf7} Performance of the QTR G-Ext method compared with the G-Ext method and XLBO algorithm. Average $\overline{k}$ and standard deviation $\sigma$ of SCF iterations. Convergence threshold $10^{-7}$.}
\begin{ruledtabular}
\begin{tabular}{lcccccccc}
      & \multicolumn{2}{c}{DMABN} & \multicolumn{2}{c}{3HF} & \multicolumn{2}{c}{AppA} & \multicolumn{2}{c}{OCP} \\
      & $\overline{k}$ & $\sigma$ & $\overline{k}$ & $\sigma$ & $\overline{k}$ & $\sigma$ & $\overline{k}$ & $\sigma$ \\ \hline
QTR G-Ext &  5.42   &  0.69  &  5.42   &  0.80  & 5.37 & 0.84 &  4.86   &  0.83  \\
G-Ext &  7.33   &  0.63  &  6.96   &  0.79  & 6.56 & 0.75 &  5.83   &  0.87 \\
XLBO  &  7.51   &  0.65  &  7.45   &  0.65  & 7.43 & 0.80  &  7.21   &  0.75  \\ 
\end{tabular}
\end{ruledtabular}
\end{table}

In conclusion, we presented the novel Quasi Time-Reversible Grassmann extrapolation scheme aimed at preserving the energy conservation of Newton's equations and, at the same time, at keeping low the number of SCF iterations. This scheme is based on the same properties of differential geometry of our previous extrapolation approach, ensuring that our guess density matrices retain all the mathematical properties of a density matrix. The innovation of this contribution lies in the symmetric combination of vectors in the tangent space, which proved to effectively preserve the stability of the total energy during the simulation. To validate its effectiveness, we conducted tests on systems of different sizes, and we obtained excellent results for all of them.

\begin{acknowledgments}

This work was supported by the Italian Ministry of University and Research under grant 2020HTSXMA\_002 (PSI-MOVIE) and by the French ‘Investissements d’Avenir’ program, project Agence Nationale de la Recherche (ISITE-BFC) (contract ANR-15-IDEX-0003). \'EP also acknowledges support from the European Research Council (ERC) under the European Union’s Horizon 2020 research and innovation program (grant agreement No 810367–project EMC2) as well as from the Simons Targeted Grant Award No. 896630. 
BS acknowledges funding by Deutsche Forschungsgemeinschaft (DFG, German Research Foundation) under Germany's Excellence Strategy – EXC 2075 – 390740016. FP is member of the GNCS group of INdAM.
\end{acknowledgments}


\bibliography{biblio}

\end{document}